\documentclass[aip,reprint]{revtex4-1}
\usepackage{graphicx}

\begin{document}

\title{Pre-earthquake magnetic pulses}

\author{John Scoville}
\affiliation{San Jose State University, Dept. of Physics, San Jose, CA 95192-0106, USA}
\affiliation{SETI Institute, Mountain View, CA 94043, USA}
\author{Jorge Heraud}
\affiliation{Pontificia Universidad Cat\'{o}lica del Per\'{u}, Lima, Peru}
\author{Friedemann Freund}
\affiliation{San Jose State University, Dept. of Physics, San Jose, CA 95192-0106, USA}
\affiliation{SETI Institute, Mountain View, CA 94043, USA}
\affiliation{NASA Ames Research Center, Earth Sci. Div. SGE, Moffett Field, CA 94035, USA}

\begin{abstract}
A semiconductor model of rocks is shown to describe unipolar magnetic pulses, a phenomenon that has been observed prior to earthquakes.  These pulses are observable because their extremely long wavelength allows them to pass through the Earth's crust.  Interestingly, the source of these pulses may be triangulated to pinpoint locations where stress is building deep within the crust.  We couple a semiconductor drift-diffusion model to a magnetic field in order to describe the electromagnetic effects associated with electrical currents flowing within rocks.  The resulting system of equations is solved numerically and it is seen that a volume of rock may act as a diode that produces transient currents when it switches bias.  These unidirectional currents are expected to produce transient unipolar magnetic pulses similar in form, amplitude, and duration to those observed before earthquakes, and this suggests that the pulses could be the result of geophysical semiconductor processes.
\end{abstract}

\maketitle

\section{Introduction}
Rocks, especially igneous rocks, behave as semiconductors under certain conditions \cite{FFchargegen, FFunified, FreundTakeuchiLau, King1984}.  Although the magnetic fields produced by small semiconductors are often negligible, semiconductors on geophysical scales may produce significant magnetic fields.  This is of particular interest since these fields can be observed at the Earth's surface and they seem to indicate that rock is being stressed deep in the crust.

Ultra-low\footnote{In this context, 'ultra-low' refers to electromagnetic waves having frequencies from millihertz to a few Hertz, in contrast to the International Telecommunications Union (ITU) definition of ultra-low, which would correspond to waves having frequencies of 300Hz-3kHz.} frequency (ULF) electromagnetic emissions have been observed prior to earthquakes \cite{Bleier2009, sharma2011,hayakawa2011}, possibly resulting from electric currents flowing deep in the crust \cite{bortnik2010}.  Low frequency fluctuations in the Earth's magnetic field have been reported in the literature since at least 1635 \cite{FraserSmithEos}.  Increased levels of magnetic fluctuations have been repeatedly observed prior to earthquakes since at least 1964 \cite{Moore1964}, but these transient phenomena have received relatively little attention and, as such, they are not yet fully understood.  

Observed pre-earthquake electromagnetic waves typically have frequencies between 0.01Hz and 20Hz, possibly owing to the fact that only low-frequency components may traverse tens of kilometers through the rock column.  They have been independently observed prior to many earthquakes over the past 50 years \cite{FraserSmithEos, Moore1964, Kopytenko, FraserSmith1990, Hayakawa1993, FraserSmith1994, Bleier2009, HeraudAGU2013}.  During the weeks leading up to the M=5.4 Alum Rock earthquake of Oct. 30, 2007, a magnetometer located about 2 km from the epicenter recorded unusual non-alternating magnetic pulses, reaching amplitudes up to 30 nT \cite{bortnik2010}. The incidence of these pulses increased as the day of the earthquake approached.  A pair of magnetometer stations in Peru recently recorded similar unipolar pulses prior to several medium-sized earthquakes, and triangulating the source of these pulses revealed the location of subsequent earthquake epicenters \cite{HeraudAGU2013}.

The unipolar magnetic pulses observed prior to earthquakes have a characteristic shape similar to a Gaussian that is attenuated over time.  The unipolar nature of the magnetic pulses is somewhat unusual and bears resemblance to pulses produced by lightning and other electrical breakdown phenomena.  However, the duration of many pre-earthquake pulses exceeds several seconds, much longer than any lightning strike.  Moreover, triangulation of such pulses near Lima, Peru revealed that strong pulses originated almost exclusively from locations within a few kilometers of future earthquake epicenters \cite{HeraudAGU2013}.

To model the electromagnetic phenomena associated with volumes of rock, we solve a three-dimensional drift-diffusion model of a semiconductor and calculate the magnetic fields induced by its electric currents.  The model is seen to describe transient low-frequency unipolar magnetic pulses. 

\section{Rocks as Semiconductors}
We will show that unipolar pulses can emerge simply from the electrical drift and random diffusion of charge carriers in a semiconducting volume of rock.  There are several reasons why this is a plausible mechanism for the observed pulses.  Large electrical currents are known to accompany earthquakes, occasionally so large that luminous effects known as earthquake lights\cite{Theriault2014} become apparent.  There is experimental evidence \cite{FFchargegen, FFunified, FreundTakeuchiLau, King1984} indicating that, during stressing, electrons and holes become activated in igneous rocks that subsequently behave as semiconductors.  

One proposed source of charge carriers in rock is the break-up of peroxy defects \cite{FFchargegen, FFunified, FreundTakeuchiLau} as a result of the increase in tectonic stresses.  The oxygen sublattice of a wide variety of silicate minerals can form peroxy defects that act as sources of electron/hole pairs \cite{FFunified}, causing these minerals to exhibit semiconductivity.  Once activated, highly mobile electronic charge carriers diffuse through the minerals.  

Peroxy defects are point defects, typically introduced through the incorporation of $\mathrm{H_2O}$ into nominally anhydrous minerals that crystallize in $\mathrm{H_2O}$-laden magmas or recrystallize in high-temperature $\mathrm{H_2O}$-laden environments\cite{FFunified}.  The incorporation of $\mathrm{H_2O}$ into oxides and silicates leads to $\mathrm{OH^{-}}$ pairs that subsequently undergo redox conversion.  The two $\mathrm{H^{+}}$ of the $\mathrm{OH^{-}}$ pairs combine to form $\mathrm{H_2}$, and the $\mathrm{O^{-}}$ ions bind to form a peroxy bond.  The formation of these peroxy bonds has been extensively studied in laboratory experiments \cite{FFunified, FFchargegen, FreundFusedSilica, Griscom2011} and treated by computational chemistry \cite{Ricci2001}.  

When peroxy bonds are energized via stresses in the rock or by heat, they may produce electron-hole pairs.  The peroxy bond breaks, forming a transient state with two unpaired electrons.  This is followed by a fully dissociated state in which a hole is free to move through the crystal structure.  A neighboring oxygen atom donates an electron and becomes a hole, as its valence shell becomes deficient by one electron.  The donated electron becomes trapped near the broken peroxy bond\cite{Griscom2011} in a new state whose energy level is slightly below the upper edge of the valence band.  In terms of the valence state, the neighboring oxygen atom, which was previously in an $\mathrm{O^{2-}}$ state, becomes $\mathrm{O^-}$.  This oxygen anion in the 1- state is effectively a positive hole with an incomplete valence shell and could also be regarded as an unstable oxygen radical \cite{FFchargegen}.  

Holes are capable of propagating through the oxygen lattice, exchanging valence electrons by a phonon-assisted vacancy hopping mechanism \cite{Shluger1992}.  This process effectively constitutes a diffusion of $\mathrm{O^-}$ holes through a lattice of $\mathrm{O^{2-}}$ atoms.  The trapped electrons are immobile but participate through recombination and electrostatic interactions.

\section{Drift-Diffusion Semiconductor Model}
The drift-diffusion equations are the most frequently used model for semiconductor physics, and perform well on scales greater than about 0.5 micrometers \cite{Vasileska}.  They describe current in terms of charge carrier concentrations and an electrostatic field, and this determines the change in charge carrier concentrations via continuity of the current density.  The drift-diffusion equations are:
\begin{eqnarray}
\partial_t \mathbf{n} &=& -\mathbf{R(n,p)} + \mathbf{\nabla} \cdot (D_n \mathbf{\nabla} \mathbf{n} - \mu_n \mathbf{n} \mathbf{\nabla} \mathbf{V}) \nonumber \\
\partial_t \mathbf{p} &=& -\mathbf{R(n,p)} + \mathbf{\nabla} \cdot (D_p \mathbf{\nabla} \mathbf{p} + \mu_p \mathbf{p} \mathbf{\nabla} \mathbf{V}) \nonumber \\
\Delta \mathbf{V} &=& \frac{1}{\epsilon} (\mathbf{n}-\mathbf{p}-\mathbf{C})
\end {eqnarray}

Here, $\mathbf{n}$, $\mathbf{p}$, $\mathbf{R}$, $\mathbf{V}$, and $\mathbf{C}$ are defined on a domain $\Omega \times (0,T)$, where $\Omega$ is a subset of a 3-dimensional space.  The functions $\mathbf{n}$ and $\mathbf{p}$ are concentrations of electron and hole charges, respectively, and $\mathbf{C}$ is the charge of any dopant ions that are present. $\mathbf{R(n,p)}$ is the recombination/generation rate of electrons and holes.  The third equation is Poisson's law of electrostatics whose solution describes the electric potential $\mathbf{V}$.  $\epsilon$ is the electric permittivity.  The constants $\mu_n$ and $\mu_p$ are the mobilities of electrons and holes, respectively, (not to be confused with the magnetic permeability $\mu$ or $\mu_0$) and $D_n$ and $D_p$ are the corresponding diffusion coefficients.  In the particular instance of the model under consideration, $\mu_n$ and $D_n$ are approximately zero due to electrons becoming trapped in the valence band.   

\section{Coupling Electromagnetism to Drift-Diffusion}
Maxwell's equations describe propagation at the speed of light, much faster than the charge carriers diffusing in a typical semiconductor.  Rather than modeling propagation on two very different time scales, we make use of the quasi-static (magnetostatic) approximation\cite{Jackson}, assuming that currents do not alternate rapidly or approach the speed of light.  Specifically, the Maxwell displacement current appearing in Ampere's law is assumed to be negligible: $c^{-2} \partial_t \mathbf{E} \approx 0$.  This assumption is implicit in the drift-diffusion model due to its use of Poisson's equation for the static electrical potential.

The electric current density $\mathbf{J}(\mathbf{x'})$ acts as the source of a magnetic field.  It may be expressed as the sum of a drift term, involving the electric field, and a diffusive term, involving the concentration gradient.  The rate of change of the concentration then becomes a continuity equation that is a function of current density.  Explicitly separating the current and continuity equations facilitates coupling to the magnetic field.  In this form, the current densities are:
\begin{eqnarray}
\mathbf{J}_n &=& D_n \nabla \mathbf{n} + \mu_n \mathbf{n} \nabla \mathbf{V} \nonumber \\
\mathbf{J}_p &=& -D_p \nabla \mathbf{p} + \mu_p \mathbf{p} \nabla \mathbf{V}
\end{eqnarray}

The continuity equations that describe the change in electron and hole concentrations are, then:
\begin{eqnarray}
\partial_t \mathbf{n} &=& -\mathbf{R(n,p)} + \nabla \cdot \mathbf{J}_n \nonumber \\
\partial_t \mathbf{p} &=& -\mathbf{R(n,p)} + \nabla \cdot (-\mathbf{J}_p)
\end{eqnarray}

The current densities $\mathbf{J}_n$ and $\mathbf{J}_p$ are summed to obtain the total current density $\mathbf{J}$ that acts as a source for the magnetic field.  In a magnetostatic approximation, the solution to the magnetic field on a domain may be efficiently computed by solving a set of Poisson equations for the magnetic vector potential.  In this case, however, we calculate the field at an arbitrary point in space, which could be outside the domain.  We apply the Biot-Savart law to obtain the magnetic field at the point $\mathbf{x}$:
\begin{equation}
\mathbf{B}(\mathbf{x},t) = \frac{\mu}{4\pi} \int (\mathbf{J}_p(\mathbf{x'},t)+\mathbf{J}_n(\mathbf{x'},t)) \times \frac {\mathbf{x}-\mathbf{x'}}{|\mathbf{x}-\mathbf{x'}|^3} d^3\mathbf{x'}
\end{equation}

Here, $|\mathbf{x}-\mathbf{x'}|$ is the magnitude of the vector from $\mathbf{x}$ to $\mathbf{x'}$ and $\mu$ is the magnetic permeability.  The velocities of the holes are not sufficiently large for the Lorentz force to be significantly influenced by magnetic fields so we do not conisder the effect of the magnetic field on the charge carriers.

\section{Numerical Solution}
The drift-diffusion equations are solved by expressing the partial differential equations as a system of ordinary differential equations for the time derivatives $\partial_t \mathbf{n}$ and $\partial_t \mathbf{p}$.  A finite-difference approximation to this system is then integrated using a fourth-order Runge-Kutta scheme (RK4).  Poisson's equation is solved separately at each timestep using successive over-relaxation\cite{GolubVanLoan} (SOR) with an adaptive relaxation parameter and open boundary conditions.  For the other PDEs, the Dirichlet boundary conditions $\mathbf{n}=0$ and $\mathbf{p}=0$ are applied to the boundary of a grid of uniformly spaced points representing the $x$, $y$, and $z$ coordinates over which functions are evaluated.  All spatial partial derivatives ($\partial_x$, $\partial_y$, $\partial_z$, $\nabla$) of the current and continuity equations are approximated using a fourth-order central difference approximation.  

At each timestep, the electric potential is determined by solving Poisson's equation, starting the SOR iteration with the electric potential from the previous timestep.  Using the electric potential and the charge carrier concentrations, the components of the current vector fields $\mathbf{J}_n$ and $\mathbf{J}_p$ are evaluated.  From $\mathbf{J}_n$ and $\mathbf{J}_p$, the continuity equations are integrated, yielding the concentrations of the charge carriers at the next timestep.  

The magnetic field $\mathbf{B}$ is evaluated by applying a discretized Biot-Savart law to the currents.  $\mathbf{B}$ is calculated at each timestep but since the result does not affect the dynamics, it may be evaluated at a single point.

\section{Results}
Since holes are mobile and electrons are immobile, diffusion separates the two species, creating an electric current that acts as an electromagnet.  The boundary of a region of activated charge behaves, essentially, like the p-n junction of a diode.  Since only holes may flow out of this volume, the initial current diffusing across the boundary is unidirectional, corresponding to forward bias in the diode.  However, after a delay period during which recombination reduces the diffusive current, the diode switches to reverse bias.  The p-n junction capacitance generates a reverse recovery current and potentially a reverse-bias electrical breakdown.  This current of holes flows back into the source volume, producing a magnetic pulse that is opposite in polarity and potentially much stronger than the initial magnetic field.

We use the semiconductor model to calculate an example of a unipolar magnetic pulse.  The electric permittivity and magnetic permeability are estimated based on the static properties of MgO\cite{Batllo1991} ($\epsilon \approx 16.75 \epsilon_0$ and $\mu \approx \mu_0$) and a temperature of $T=673.15$K.  Since electrons are trapped and immobile in broken peroxy bonds, $\mu_n$ and $D_n$ are set to $0$.  The mobility and diffusion constant of holes were roughly estimated based on experimental data from a Haynes-Shockley-style experiment whereby a rapid pressure impulse to the center of a gabbro tile injected holes that diffused and drifted away from their source.  The parameters used are $\mu_p= 0.063$ m$^2$ (Vs)$^{-1}$ and $D_p=8.5 \times 10^{-4}$ m$^2$ s$^{-1}$, comparable to their values in pure undoped silicon.  

Charge generation is not explicitly considered in this calculation, and a pre-existing excess concentration of $10^{-5}$ C m$^{-3}$ of both electrons and holes exists as an initial condition.  These dissociated charges are initially present only at points within a ball of radius $10^3$ m.  Their recombination rate is proportional to the product of electron and hole concentrations, $\mathbf{R} = 10^{22} \mathbf{np}$.  The attenuation of the magnetic field as it passes through the Earth is not considered, nor are effects associated with the surface of the Earth.  

\begin{figure}[placement h]
\centering
  \includegraphics[width=3.3in]{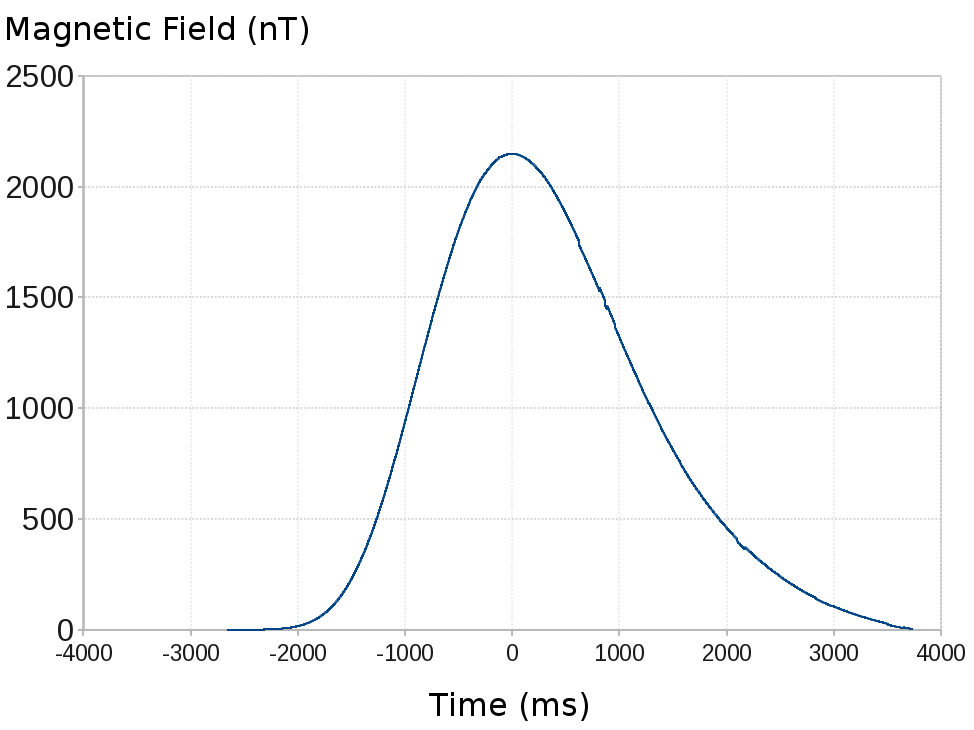}
  \includegraphics[width=3.3in]{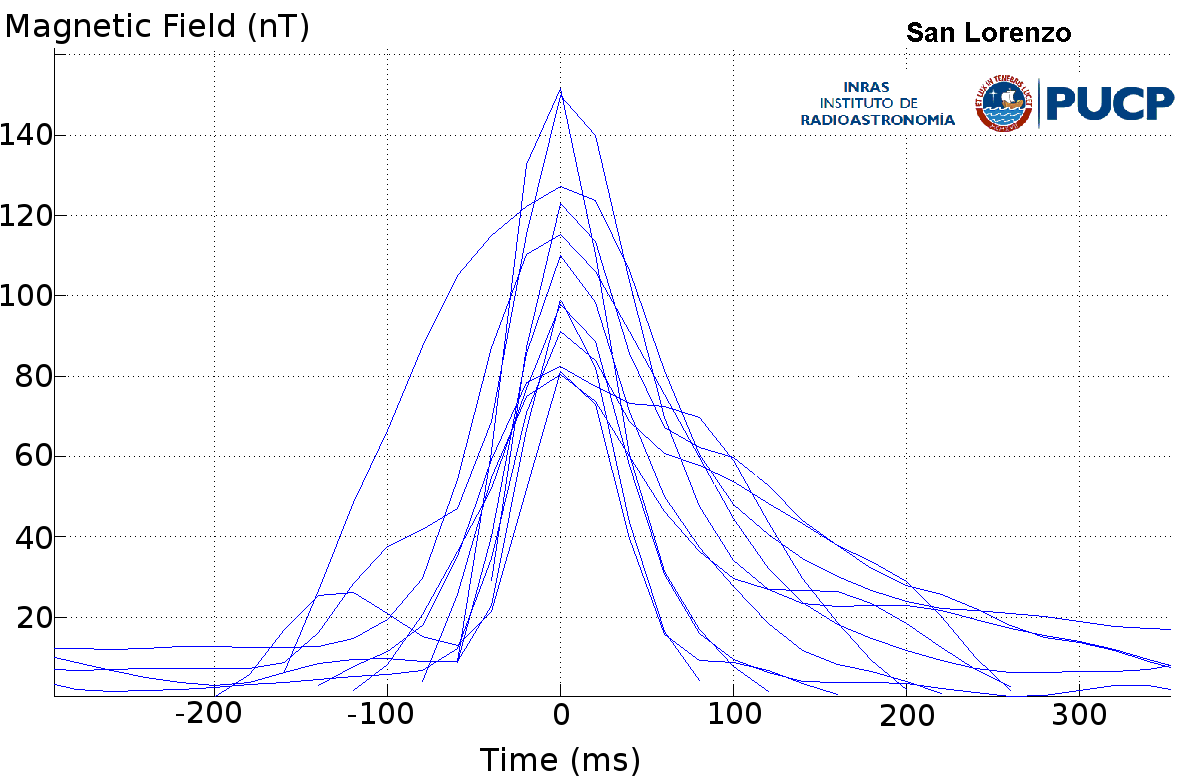}
  \caption{(Color online) Above, a calculated transient magnetic pulse, 10km from the current source.  Below, pulses observed prior to an earthquake in Lima, Peru, approximately 25km from the epicenter.}
\end{figure}

Calculated and observed magnetic pulses are illustrated in Figure 1.  The first figure shows the value of the x-component of the magnetic field as a function of time, measured 10km directly above the center of the simulated volume.  The amplitude, frequency, and shape of the pulse are similar to pulses that have been observed before earthquakes.  For comparison, the second figure shows several magnetic pulses observed prior to an earthquake near Lima, Peru.  These pulses were measured over a period of several days by a pair of magnetometers approximately 25km away, and the locations of their sources were triangulated.  The sources were clustered within a few kilometers of the epicenter of an earthquake that occurred two weeks after the onset of the pulses \cite{HeraudAGU2013}.  This analysis has been performed prior to several moderate earthquakes near Lima, with similar results.

\section{Conclusion}
\indent When a volume of rock is stressed, excess holes and electrons are injected.  The mobile holes begin to diffuse out of the source volume, while electrons are trapped within the source volume and undergo recombination with the holes that have not diffused out.  The flux of holes leaving the source effectively creates a p-n diode: the source volume becomes an n-type semiconductor and the surrounding rock becomes p-type.  A depletion region forms between the two layers of the p-n junction and the p-n double layer screens electric fields outside its immediate vicinity.

After charge injection, a diffusive current of holes flows as a result of the concentration gradient across the source boundary.  This corresponds to a forward bias state of the diode, dominated by diffusion capacitance rather than junction capacitance.  This current creates a transient magnetic field.  As the hole concentration gradient decreases, the diffusive current and the magnetic field decay.  After holes have diffused outward, creating p-type and n-type regions, a junction capacitance results from layers of positive and negative charge separated by a depletion region at the junction.  

After a delay period, the diode effectively switches to a reverse bias state.  Electron-hole recombination consumes the holes remaining within the source volume, leaving mostly electrons inside.  The junction capacitance causes a transient reverse recovery current, and, if the potential drop across the depletion region is sufficiently strong, reverse-bias electrical breakdown may occur.  Coulomb attraction pulls the holes back into the source volume, and this reverse current of holes can create transient magnetic pulses similar to those observed before earthquakes.

A distinctive form and the ability to pass through the earth at ultra-low frequencies make magnetic pulses a compelling tool for the observation of pre-seismic shifts in the stress level of rocks that are otherwise inaccessible due to depth.  By triangulating the source of these magnetic pulses, the increased buildup of stress around future earthquake epicenters may be identified weeks in advance of seismicity.

In addition to unipolar pulses, other types of electromagnetic precursors might be predicted from the semiconductor model.  Oscillatory ULF fields, for example, have been observed immediately preceding earthquake activity \cite{Bleier2009}.  

The 'positive hole' semiconductor theory modeled here seeks to unify a wide range of electromagnetic phenomena associated with seismic activity.  The direct coupling of semiconductor drift-diffusion currents and electromagnetism produces a model consistent with observations of pre-seismic magnetic pulses.  This suggests that pre-earthquake ULF activity may be the result of geophysical semiconductor processes.

\begin{acknowledgments}
\indent The authors would like to thank Tom Bleier and Clark Dunson of QuakeFinder for informative discussions.  This research was funded in part by NASA Earth Surface and Interior Grant NNX12AL71G (Dr. John LaBrecque).
\end{acknowledgments}

\bibliographystyle{unsrt}

\end{document}